\documentstyle[prl,aps]{revtex}

\begin{document}
\author{Jian-Qi Shen $^{1,}$$^{2}$ \footnote{E-mail address: jqshen@coer.zju.edu.cn}}
\address{$^{1}$  Centre for Optical
and Electromagnetic Research, State Key Laboratory of Modern
Optical Instrumentation\\
$^{2}$Zhejiang Institute of Modern Physics and Department of
Physics\\
Zhejiang University, Spring  Jade, Hangzhou 310027, P.R. China}
\date{\today }
\title{Wave Propagation in Generalized Gyrotropic Media\footnote{It will be submitted nowhere else for the publication, just uploaded at the e-print archives.}}
\maketitle
\begin{abstract}
Wave propagation inside generalized gyrotropic media is considered
in the present Letter. It is shown that in this type of materials,
the left- and right- handed circularly polarized light can be
coupled to each other, which may possess some potential
applications to quantum information.
\\ \\
\end{abstract}
Gyrotropic media whose electric permittivity and magnetic
permeability are tensors has peculiar optical and electromagnetic
properties\cite{Veselago,Li}. Recently we proposed an
experimentally feasible scheme of testing the so-called quantum
vacuum geometric phases of zero-point electromagnetic fields in
the helically curved optical fibre by using the gyrotropic-medium
fibre\cite{Shen}. In this Letter, we investigate the wave
propagation in a generalized gyrotropic medium with the following
permittivity and permeability
\begin{equation}
(\epsilon)_{ik}=\left(\begin{array}{cccc}
a_{\rm e}+d_{\rm e}  & b_{\rm e}-ic_{\rm e} & 0 \\
b_{\rm e}+ic_{\rm e} & a_{\rm e}-d_{\rm e} & 0  \\
 0 &  0 &  \epsilon_{3}
 \end{array}
 \right),         \quad    (\mu)_{ik}=\left(\begin{array}{cccc}
a_{\rm h}+d_{\rm h}  & b_{\rm h}-ic_{\rm h} & 0 \\
b_{\rm h}+ic_{\rm h} & a_{\rm h}-d_{\rm h} & 0  \\
 0 &  0 &  \mu_{3}
 \end{array}
 \right).        \label{eq1}
\end{equation}
This type of materials is a generalization of the ordinary
gyrotropic medium with\cite{Veselago,Li}
\begin{equation}
(\epsilon)_{ik}=\left(\begin{array}{cccc}
\epsilon_{1}  & i\epsilon_{2} & 0 \\
-i\epsilon_{2} &   \epsilon_{1} & 0  \\
 0 &  0 &  \epsilon_{3}
 \end{array}
 \right),                 \qquad          (\mu)_{ik}=\left(\begin{array}{cccc}
\mu_{1}  & i\mu_{2} & 0 \\
-i\mu_{2} &   \mu_{1} & 0  \\
 0 &  0 &  \mu_{3}
 \end{array}
 \right)              \label{eq2}
\end{equation}
and can encompass the electromagnetic parameters of
$\epsilon_{ik}$ and $\mu_{ik}$ in (\ref{eq1}) material responses
experimentally obtained.

Assume that the wave vector ${\bf k}$ of electromagnetic wave
propagating inside this medium is parallel to the the third
component of the Cartesian coordinate system. According to
Maxwellian Equations, one can arrive at
\begin{equation}
-\nabla^{2}{
E}_{1}=\mu_{0}\left(\mu_{21}\frac{\partial^{2}}{\partial
t^{2}}D_{2}-\mu_{22}\frac{\partial^{2}}{\partial
t^{2}}D_{1}\right),   \quad   -\nabla^{2}{
E}_{2}=-\mu_{0}\left(\mu_{11}\frac{\partial^{2}}{\partial
t^{2}}D_{2}-\mu_{12}\frac{\partial^{2}}{\partial
t^{2}}D_{1}\right),
\end{equation}
where the electric displacement vector $D_{1}$ and $D_{2}$ are of
the form
\begin{equation}
D_{1}=\epsilon_{0}\left(\epsilon_{11}E_{1}+\epsilon_{12}E_{2}\right),
\quad
D_{2}=\epsilon_{0}\left(\epsilon_{21}E_{1}+\epsilon_{22}E_{2}\right).
\end{equation}
Due to the transverse nature of planar electromagnetic wave, the
following acquirements are satisfied
\begin{equation}
E_{3}=0,\quad  H_{3}=0, \quad  {\bf k}\cdot{\bf E}=0,  \quad  {\bf
k}\cdot{\bf H}=0,  \quad  {\bf k}\cdot{\bf D}=0,    \quad  {\bf
k}\cdot{\bf B}=0.
\end{equation}
Thus with the help of above equations, it is verified that
\begin{eqnarray}
\nabla^{2}{
E}_{1}=-\epsilon_{0}\mu_{0}\left[\left(\mu_{21}\epsilon_{21}-\mu_{22}\epsilon_{11}\right)\frac{\partial^{2}}{\partial
t^{2}}E_{1}+\left(\mu_{21}\epsilon_{22}-\mu_{22}\epsilon_{12}\right)\frac{\partial^{2}}{\partial
t^{2}}E_{2}\right],
\nonumber \\
\nabla^{2}{
E}_{2}=\epsilon_{0}\mu_{0}\left[\left(\mu_{11}\epsilon_{21}-\mu_{12}\epsilon_{11}\right)\frac{\partial^{2}}{\partial
t^{2}}E_{1}+\left(\mu_{11}\epsilon_{22}-\mu_{12}\epsilon_{12}\right)\frac{\partial^{2}}{\partial
t^{2}}E_{2}\right].
\end{eqnarray}
So,
\begin{eqnarray}
\nabla^{2}\left(\frac{E_{1}\pm
iE_{2}}{\sqrt{2}}\right)&=&\frac{1}{\sqrt{2}}\epsilon_{0}\mu_{0}\left[\left(\mu_{22}\epsilon_{11}-\mu_{21}\epsilon_{21}\right)\pm
i\left(\mu_{11}\epsilon_{21}-\mu_{12}\epsilon_{11}\right)\right]\frac{\partial^{2}}{\partial
t^{2}}E_{1}           \nonumber \\
&+&\frac{1}{\sqrt{2}}\epsilon_{0}\mu_{0}\left[\left(\mu_{22}\epsilon_{12}-\mu_{21}\epsilon_{22}\right)\pm
i\left(\mu_{11}\epsilon_{22}-\mu_{12}\epsilon_{12}\right)\right]\frac{\partial^{2}}{\partial
t^{2}}E_{2}.
\end{eqnarray}
It follows from (\ref{eq1}) that
\begin{eqnarray}
\left(\mu_{22}\epsilon_{11}-\mu_{21}\epsilon_{21}\right)+
i\left(\mu_{11}\epsilon_{21}-\mu_{12}\epsilon_{11}\right)=A_{+}+iB_{+},
 \nonumber \\
\left(\mu_{22}\epsilon_{12}-\mu_{21}\epsilon_{22}\right)+
i\left(\mu_{11}\epsilon_{22}-\mu_{12}\epsilon_{12}\right)=iA_{+}+B_{+},
\end{eqnarray}
where
\begin{eqnarray}
A_{+}=a_{\rm e}a_{\rm h}-a_{\rm e}c_{\rm h}-d_{\rm e}d_{\rm
h}-id_{\rm e}b_{\rm h}-b_{\rm e}b_{\rm h}+ib_{\rm e}d_{\rm
h}+c_{\rm e}c_{\rm h}-c_{\rm e}a_{\rm h},
 \nonumber \\
 B_{+}=-a_{\rm e}b_{\rm h}-id_{\rm e}a_{\rm h}+id_{\rm e}c_{\rm
 h}-b_{\rm e}c_{\rm h}+b_{\rm e}a_{\rm h}-c_{\rm e}b_{\rm h}+ic_{\rm e}d_{\rm
 h}+ia_{\rm e}d_{\rm h}.
\end{eqnarray}
In the same fashion, one can arrive at
\begin{eqnarray}
\left(\mu_{22}\epsilon_{11}-\mu_{21}\epsilon_{21}\right)-
i\left(\mu_{11}\epsilon_{21}-\mu_{12}\epsilon_{11}\right)=A_{-}+iB_{-},
 \nonumber \\
 \left(\mu_{22}\epsilon_{12}-\mu_{21}\epsilon_{22}\right)-
i\left(\mu_{11}\epsilon_{22}-\mu_{12}\epsilon_{12}\right)=-iA_{-}-B_{-},
\end{eqnarray}
where
\begin{eqnarray}
A_{-}=a_{\rm e}a_{\rm h}+a_{\rm e}c_{\rm h}-d_{\rm e}d_{\rm
h}+id_{\rm e}b_{\rm h}-b_{\rm e}b_{\rm h}-ib_{\rm e}d_{\rm
h}+c_{\rm e}c_{\rm h}+c_{\rm e}a_{\rm h},
 \nonumber \\
 B_{-}=a_{\rm e}b_{\rm h}-id_{\rm e}a_{\rm h}-id_{\rm e}c_{\rm
 h}-b_{\rm e}c_{\rm h}-b_{\rm e}a_{\rm h}-c_{\rm e}b_{\rm h}-ic_{\rm e}d_{\rm
 h}+ia_{\rm e}d_{\rm h}.
\end{eqnarray}
Hence, the wave equations of left- and right- handed circularly
polarized light\footnote{For simplicity, without loss of
generality, it is assumed that the two mutually perpendicular real
unit polarization vectors ${\vec\varepsilon}(k,1)$ and
${\vec\varepsilon}(k,2)$ can be taken to be as follows:
$\varepsilon_{1}(k,1)=\varepsilon_{2}(k,2)=1$,
$\varepsilon_{1}(k,2)=\varepsilon_{2}(k,1)=0$ and
$\varepsilon_{3}(k,1)=\varepsilon_{3}(k,2)=0$. Thus by the aid of
the formula ${\bf E}=-\frac{\partial {\bf A}}{\partial t}$ for the
electric field strength, in the second-quantization framework one
can arrive at\cite{Shen}
\begin{eqnarray}
E_{R}=\frac{E_{1}+ iE_{2}}{\sqrt{2}}=i\int{\rm d}^{3}{\bf
k}\sqrt{\frac{\omega}{2(2\pi)^{3}}}[a_{L}(k)\exp(-ik\cdot
x)-a_{R}^{\dagger}(k)\exp(ik\cdot x)],   \nonumber  \\
E_{L}=\frac{E_{1}-iE_{2}}{\sqrt{2}}=i\int{\rm d}^{3}{\bf
k}\sqrt{\frac{\omega}{2(2\pi)^{3}}}[a_{R}(k)\exp(-ik\cdot
x)-a_{L}^{\dagger}(k)\exp(ik\cdot x)].               
\end{eqnarray} } is written
\begin{eqnarray}
 \nabla^{2}\left(\frac{E_{1}-
iE_{2}}{\sqrt{2}}\right)=\epsilon_{0}\mu_{0}A_{-}\frac{\partial^{2}}{\partial
t^{2}}\left(\frac{E_{1}-
iE_{2}}{\sqrt{2}}\right)+i\epsilon_{0}\mu_{0}B_{-}\frac{\partial^{2}}{\partial
t^{2}}\left(\frac{E_{1}+iE_{2}}{\sqrt{2}}\right),
 \nonumber \\
 \nabla^{2}\left(\frac{E_{1}+
iE_{2}}{\sqrt{2}}\right)=\epsilon_{0}\mu_{0}A_{+}\frac{\partial^{2}}{\partial
t^{2}}\left(\frac{E_{1}+
iE_{2}}{\sqrt{2}}\right)+i\epsilon_{0}\mu_{0}B_{+}\frac{\partial^{2}}{\partial
t^{2}}\left(\frac{E_{1}-iE_{2}}{\sqrt{2}}\right).
\label{eq13}
\end{eqnarray}

It should be pointed out that the interaction between left- and
right-handed polarized light arises in Eq.(\ref{eq13} ), which is
analogous to the Josephson's effect and may therefore be of
physical interest.

The coupling of left-handed polarized light to the right-handed
one can be characterized by the frequency shift $\Omega_{L}$ and
$\Omega_{R}$, namely, the amplitudes of left- and right- handed
circularly polarized light propagating along the
$\hat{z}$-direction are written as follows
\begin{equation}
E_{L}\sim\exp\left\{\frac{1}{i}\left[\left(\omega+\Omega_{L}\right)t-\sqrt{A_{-}}\frac{\omega}{c}
z+\phi_{L}\right]\right\},  \quad
E_{R}\sim\exp\left\{\frac{1}{i}\left[\left(\omega+\Omega_{R}\right)t-\sqrt{A_{+}}\frac{\omega}{c}
z+\phi_{R}\right]\right\}.     \label{eq15}
\end{equation}
Insertion of Eq.(\ref{eq15}) into Eq.(\ref{eq13}) yields
 \begin{equation}
A_{-}\omega^{2}E_{L}=A_{-}\left(\omega+\Omega_{L}\right)^{2}E_{L}+iB_{-}\left(\omega+\Omega_{R}\right)^{2}E_{R},
\quad
A_{+}\omega^{2}E_{R}=A_{+}\left(\omega+\Omega_{R}\right)^{2}E_{R}+iB_{+}\left(\omega+\Omega_{L}\right)^{2}E_{L},
     \label{eq16}
\end{equation}
and consequently
 \begin{equation}
A_{+}A_{-}\left(2\omega\Omega_{R}+\Omega_{R}^{2}\right)\left(2\omega\Omega_{L}+\Omega_{L}^{2}\right)+B_{+}B_{-}\left(\omega+\Omega_{L}\right)^{2}\left(\omega+\Omega_{R}\right)^{2}=0,
     \label{eq17}
\end{equation}
which is a restricted condition regarding the frequency-shift
relation between left- and right- handed circularly polarized
light.

Note that in the case of conventional gyrotropic media
characteristic of such permittivity and permeability tensors
(\ref{eq2}), where $d_{\rm e}=b_{\rm e}=d_{\rm h}=b_{\rm h}=0$,
$a_{\rm e }=\epsilon_{1}$, $c_{\rm e }=-\epsilon_{2}$, $a_{\rm h
}=\mu_{1}$, $c_{\rm h}=-\mu_{2}$, the optical refractive index
squared is of the form
\begin{equation}
A_{\pm}=\left(\epsilon_{1}\pm\epsilon_{2}\right)\left(\mu_{1}\pm\mu_{2}\right),
\end{equation}
and the coupling coefficients $B_{\pm}$ are vanishing.
\\ \\

More recently, one of the artificial composite metamaterials, the
left-handed medium which has a frequency band (GHz) where the
electric permittivity ($\epsilon $) and the magnetic permeability
($\mu $) are simultaneously negative, has focused attention of
many authors both experimentally and
theoretically\cite{Smith,Klimov,Pendry3,Shelby,Ziolkowski2}. In
the left-handed medium, most phenomena as the Doppler effect,
Vavilov-Cherenkov radiation and even Snell's law are inverted. In
1964\footnote{Note that, in the literature, several authors
mention the year when Veselago suggested the {\it left-handed
media} by mistake. They claim that Veselago proposed the concept
of {\it left-handed media} in 1968. On the contrary, the true fact
is as follows: Veselago's excellent paper was first published in
July, 1964 [Usp. Fiz. Nauk {\bf 92}, 517-526 (1964)]. In 1968,
this original paper was translated into English by W. H. Furry and
published again in the journal of Sov. Phys.
Usp.\cite{Veselago}.}, Veselago first considered many peculiar
optical and electromagnetic properties, phenomena and effects in
this medium and referred to such materials as left-handed media,
since in this case the propagation vector \textbf{k}, electric
field \textbf{E} and magnetic field \textbf{H} of light wave
propagating inside it form a left-handed system\cite{Veselago}. It
follows from Maxwell's curl equations that such media having
negative simultaneously negative $\epsilon $ and $\mu $ exhibit a
negative index of
refraction, {\it i.e.}, $n=-\sqrt{\epsilon \mu }$. In experiments, the negative $%
\epsilon $ and $\mu $ can be respectively realized by using a
network (array) of thin (long) metal wires\cite{Pendry} and a
periodic arrangement of split ring resonators\cite{Pendry3}. A
combination of the two structures yields a left-handed medium.
Veselago's original paper and most of the recent theoretical works
discussed mainly the characteristics of electromagnetic wave
propagation in isotropic left-handed media, but up to now, the
left-handed media that have been prepared successfully
experimentally are actually anisotropic in nature, and it may be
very difficult to prepare an isotropic left-handed
medium\cite{Hu}. Hu and Chui presented a detailed investigation on
the characteristics of electromagnetic wave propagation in
uniaxially anisotropic left-handed media\cite{Hu}. It is believed
that the above-presented treatment of wave propagation in
generalized gyrotropic media may be applicable to the uniaxially
anisotropic left-handed materials.
\\ \\
\textbf{Acknowledgements}  I thank S. L. He for his helpful
suggestions about the wave propagation in left-handed media. This
project is supported by the National Natural Science Foundation of
China under the project No. $90101024$.

\end{document}